# Faraday spectroscopy in an optical lattice: a continuous probe of atom dynamics

*Greg Smith, Souma Chaudhury and Poul S. Jessen,*
*Optical Sciences Center, University of Arizona, Tucson, AZ 85721*

**Abstract.**
The linear Faraday effect is used to implement a continuous measurement of the spin of a sample of laser cooled atoms trapped in an optical lattice. One of the optical lattice beams serves also as a probe beam, thereby allowing one to monitor the atomic dynamics in real time and with minimal perturbation. A simple theory is developed to predict the measurement sensitivity and associated cost in terms of decoherence caused by the scattering of probe photons. Calculated signal-to-noise ratios in measurements of Larmor precession are found to agree with experimental data for a wide range of lattice intensity and detuning. Finally, quantum backaction is estimated by comparing the measurement sensitivity to spin projection noise, and shown to be insignificant in the current experiment. A continuous quantum measurement based on Faraday spectroscopy in optical lattices may open up new possibilities for the study of quantum feedback and classically chaotic quantum systems.

## 1. Introduction

It is well known that optically pumped atomic vapors can exhibit a significant Faraday effect for near-resonant probe light [1]. Even in relatively dilute samples, such as those produced by laser cooling in magneto-optic traps, the resulting polarization rotation is easily measurable and can be used as a sensitive monitor for the spin degrees of freedom [2]. Kuzmich et al. [3] and Takahashi et al. [4] have shown that the coupling between the Stokes vector of the light and the total many body spin is of the form required for a quantum non-demolition (QND) measurement of the latter. For a sufficiently strong atom-field coupling the sensitivity may be below the quantum uncertainty associated with a spin-coherent state, and may allow for the generation of spin squeezing. In a single-pass geometry this occurs when the sample is optically thick on resonance. An experimental demonstration of spin squeezing in an atomic beam has been performed by Kuzmich et al. [5] Along the same lines Julsgaard et al. measured the Faraday rotation of a probe beam passing through two separate, room temperature vapor cells and so created a modest amount of entanglement between the corresponding many body spins [6].

In this work we examine the use of Faraday spectroscopy to probe the collective spin of a sample of ultracold atoms trapped in a far-off-resonance optical lattice. Our goal is to implement a continuous, non-perturbing measurement scheme which allows to monitor coherent dynamics in the atom-lattice system. This is accomplished by measuring the polarization rotation of one of the component lattice beams as it passes through the atomic sample. We give a detailed evaluation of the sensitivity and sensitivity/decoherence tradeoff as a function of the lattice parameters, and discuss the conditions necessary to have significant measurement backaction on the atomic sample. The optical lattice used for these initial investigations is a special ($\theta = 0$) case of the 1D lin-$\theta$-lin configuration, which consists of a pair of counter propagating laser beams with linear polarizations forming an angle $\theta$. In this configuration the lattice potential is spin-independent [7], and it is straightforward to spin-polarize the atoms by optical pumping and to induce Larmor precession in an external magnetic field. Eventually we plan to study the much richer dynamics that can arise in lattices with $\theta \neq 0$ and in the presence of magnetic fields. In that case one can design an atom-lattice interaction which couples the spin and center-of-mass motion of individual atoms, so that coherent evolution in the lattice can produce spinor wavepackets with strong entanglement between spin- and space degrees of freedom. This allows the spin to serve as a meter for the entire quantum dynamics, as demonstrated in the study of atom tunneling in mesoscopic optical double wells [8]. Ghose et al. have shown that a classical model of the coupled spin-center-of-mass motion in this system exhibits deterministic chaos [9]. The prospect of using Faraday spectroscopy to implement a continuous measurement then makes the atom/lattice system a good candidate for studies of quantum chaos and – ultimately – the role played by quantum measurement in the emergence of classical chaos. Faraday spectroscopy can also provide a signal for feedback control of the collective atomic spin, and perhaps even for the more complex spin-motion dynamics in a 1D lin-$\theta$-lin lattice. If the measurement sensitivity can be increased to the point of significant backaction this will make the atom-lattice system an attractive platform for studies of quantum feedback [10].

**2. Faraday spectroscopy**

Linear magneto-optics, including the Faraday effect, have been studied since the 1800's, and the theoretical understanding of these phenomena is very mature. Here we review just a few of the salient features of the Faraday effect in alkali atom vapors, in order to estimate the measurement sensitivity and determine how it scales with experimental parameters. Faraday rotation of a linearly polarized probe field arises from different indices of refraction for left- and right-hand circular polarized light. These indices in turn depend on the diagonal elements of the polarizability tensor. A discussion of the tensor polarizability of alkali atoms in the usual regime where the probe detuning is large compared to the excited state hyperfine splitting can be found in for example ref. [7]. From the diagonal elements of the polarizability tensor in spherical coordinates we obtain

$$n_\pm = 1 + \frac{\rho \alpha(\Delta)}{2\varepsilon_0}\left(\frac{2}{3} \pm \frac{1}{3}\frac{\langle \hat{F}_z \rangle}{F}\right), \qquad (1)$$

where $\alpha(\Delta) = -3\varepsilon_0 \lambda^3 \Gamma / 8\pi^2 \Delta$ is the scalar polarizability for a two-level atom with transition wavelength $\lambda$ and natural linewidth $\Gamma$, in the large detuning, low saturation limit. Also, $\rho$ is the atom density and $\Delta = \omega - \omega_0$ is the detuning from resonance. For a linearly polarized, traveling wave probe field the differential phase shift between the $\sigma_+$ and $\sigma_-$ components is then

$$\varphi = (n_+ - n_-)kl = -\frac{1}{6}\frac{\sigma\rho l}{\Delta/\Gamma}\frac{\langle \hat{F}_z \rangle}{F}, \qquad (2)$$

where $\sigma = 3\lambda^2/2\pi$ is the resonant photon scattering cross section, $l$ is the optical path length through the sample, and $\sigma\rho l$ is the optical depth on resonance. For a probe intensity $I_P$ and input polarization $\boldsymbol{\varepsilon}_P = \boldsymbol{\varepsilon}_x \cos(\pi/4) + \boldsymbol{\varepsilon}_y \sin(\pi/4)$, the difference in output intensity for the $\boldsymbol{\varepsilon}_x$ and $\boldsymbol{\varepsilon}_y$ components is

$$\Delta I = I_x - I_y = -I_L \sin(\varphi) \approx -I_L \varphi. \qquad (3)$$

In our experiment the atomic density distribution is approximately Gaussian and contains a total number of atoms $N$ within a $1/e$ radius $L$. In that case different parts of the probe beam see different optical depths and therefore different amounts of Faraday rotation. We detect Faraday rotation by measuring the total power difference between the $\boldsymbol{\varepsilon}_x$ and $\boldsymbol{\varepsilon}_y$ components, $\Delta P_S = P_x - P_y$, in an aperture of radius $a$ centered on the atom cloud. Replacing $\rho l$ in eq. 3 with the local column density and integrating eq. 3 over the detection aperture, we obtain the result

$$\Delta P_S = \int_0^a -I_L \varphi(r) 2\pi r dr = \frac{1}{6}\frac{P}{\Delta/\Gamma}\frac{\sigma N}{\pi a^2}\left(1 - e^{-a^2/2L^2}\right)\frac{\langle \hat{F}_z \rangle}{F}, \qquad (4)$$

where $P$ is the total power passing through the aperture. Here we have assumed that the probe beam is much larger than the size of the atom cloud, so that the probe intensity is effectively constant across the aperture.

To determine the fundamental limit on sensitivity, the result of eq. 4 must be compared to the fluctuations caused by shot-noise. The shot-noise is equivalent to fluctuations in the power difference with a root-mean-square (RMS) amplitude of

$$\Delta P_N = \sqrt{\frac{P\hbar\omega}{2\kappa\tau_{pd}}}, \qquad (5)$$

where $\kappa$ is the quantum efficiency and $\tau_{pd}$ the time constant of the photo-detector [11]. The smallest detectable spin polarization is found by setting $\Delta P_S = \Delta P_N$, giving

$$\frac{\delta F_z}{F} = \sqrt{\frac{\hbar\omega_0}{2\kappa\tau_{pd}}}\frac{6\pi a^2}{\sigma N\left(1 - e^{-a^2/2L^2}\right)}\frac{\Delta/\Gamma}{\sqrt{P}}. \qquad (6)$$

Noting that the rate of photon scatting per atom is

$$\gamma_s = \tau_s^{-1} = \frac{\Gamma}{12}\frac{I/I_0}{(\Delta/\Gamma)^2} \propto \frac{P}{(\Delta/\Gamma)^2},\tag{7}$$

where the saturation intensity $I_0 = 2\pi^2\hbar c\Gamma/3\lambda^3$, we can rewrite eq. 6 as

$$\frac{\delta F_z}{F} = \frac{\sqrt{2}\pi a}{\lambda N\sqrt{\kappa}\left(1-e^{-a^2/2L^2}\right)}\sqrt{\frac{\tau_s}{\tau_{pd}}}.\tag{8}$$

This expression is minimized by setting $a/L \approx 1.58$ leading to the simple expression

$$\frac{\delta F_z}{F} \approx 9.8\frac{L}{\lambda N\sqrt{\kappa}}\sqrt{\frac{\tau_s}{\tau_{pd}}}.\tag{9}$$

For the purpose of comparison with experiment it is convenient to define another figure of merit, the signal-to-noise ratio (SNR) with which we can measure a change in $\langle\hat{F}_z\rangle$ from 0 to $F$,

$$SNR = \frac{F}{\delta F_z} \approx 0.10\frac{\lambda N\sqrt{\kappa}}{L}\sqrt{\frac{\tau_{pd}}{\tau_s}}.\tag{10}$$

Equations 9 and 10 predicts the sensitivity of our Faraday measurement, and shows how it varies with probe parameters. Notably, the dependence on probe intensity and detuning is completely contained in $\tau_s = \gamma_s^{-1}$, the mean time between photon scattering events. In a concrete experiment the detector time constant $\tau_{pd}$ determines the fastest changes that we can see, while the photon scattering time $\tau_s$ sets the timescale for decoherence. If we are trying to observe coherent dynamics on some characteristic time scale $\tau$, then we clearly need to choose probe parameters and detector bandwidth so that $\tau_{pd} < \tau < \tau_s$, say $\tau_s \sim 10\,\tau_{pd}$.

**2.1 Quantum backaction.**

It is instructive to consider when a measurement becomes sensitive enough to create significant quantum backaction onto the atomic ensemble. This problem has previously been addressed in the context of QND measurements [3,4]. Here we discuss the issue in the context of spin projection noise [12] and spin-squeezing, in order to provide a simple physical picture. Let

$$\tilde{\mathbf{F}} = \sum_{i=0}^{N}\mathbf{F}^{(i)}\tag{11}$$

be the collective spin of a sample of $N$ atoms. The quantum uncertainty in a measurement of one of the individual $F_z^{(i)}$ is given by the uncertainty relation, $\Delta F_y^{(i)2} \Delta F_z^{(i)2} \geq (\hbar/2) \left|\left\langle \Delta F_x^{(i)} \right\rangle\right|$. If we prepare a spin-coherent state aligned along the $x$-axis, then $\Delta F_y^{(i)} = \Delta F_z^{(i)} = \Delta F_z$, and $\left|\left\langle \Delta F_x^{(i)} \right\rangle\right| = \hbar F$. We then get a total uncertainty for the $z$-component of the total many body spin,

$$\Delta \tilde{F}_z^2 = N \Delta F_z^2 \Rightarrow \frac{\Delta \tilde{F}_z}{\tilde{F}} = \frac{1}{\sqrt{2F}} \frac{1}{\sqrt{N}}. \tag{12}$$

If we measure $\tilde{F}_z$ with a precision better than $\Delta \tilde{F}_z$ then the result is a spin-squeezed state. In other words, when the measurement noise is less than the spin projection noise then we gain new information about the spin statistics, the post-measurement quantum state has a more precisely known value of $\tilde{F}_z$ than the initial coherent state, and we have non-negligible backaction. Next, we note that in eq. 4 we can set $\langle F_z \rangle / F = \left\langle \sum_i F_z^i / F \right\rangle = \langle \tilde{F}_z \rangle / \tilde{F}$, which suggests that the Faraday rotation depends on the $z$-component of the total spin. This is confirmed by a more complete analysis including a quantized electromagnetic field [3,4]. Thus, in eq. 9 we can set $\delta F_z / F = \Delta \tilde{F}_z / \tilde{F}$, and the condition for non-negligible backaction becomes

$$\frac{\Delta \tilde{F}_z}{\tilde{F}} > \frac{\delta \tilde{F}_z}{\tilde{F}} \Rightarrow 0.071 \frac{\lambda}{L} \sqrt{\frac{\kappa N}{F}} \sqrt{\frac{\tau_{pd}}{\tau_s}} = \eta > 1, \tag{13}$$

where we have defined a figure of merit $\eta$ that characterizes the strength of the measurement and the significance of backaction. It is sometimes convenient to express this in terms of the optical density $O = \sigma N / 2\pi L^2$, at resonance and measured through the center of the atom cloud,

$$0.26 \sqrt{\frac{\kappa O}{F}} \sqrt{\frac{\tau_{pd}}{\tau_s}} = \eta > 1. \tag{14}$$

Equation 14 suggests that if $\tau_s \sim 10 \tau$ it might require resonant optical depths of order $10^3$ to enter the regime of significant backaction. Note, however, that we have so far considered an aperture size that maximizes the Faraday rotation signal, with the consequence that a large part of the probe field passes through the cloud away from the center where the optical depth is maximum. The numerical prefactor on left hand side of eq. 14, by contrast, is maximized for an aperture size $a \to 0$. For $a/L \leq 0.2$ it is essentially independent of $a$ and increases to $\sim 0.41$.

**2.2 Faraday rotation in an optical lattice.**
So far we have considered Faraday spectroscopy in a traveling wave geometry. In our experiment the probe field is part of an optical lattice, and the atoms are cold enough to be tightly confined and spatially ordered in the periodic lattice potential. Previous work on Bragg scattering from optical lattices has demonstrated that this spatial order is automatically of a form that allows each lattice beam to Bragg scatter in the direction of the other lattice beams [13]. The result is a few minor but nevertheless important modifications of the results derived so far.

In a Faraday active medium the atomic response includes a dipole component orthogonal to the (linear) driving polarization. It is the interference between the probe field and the "signal" field radiated by the atomic dipoles which produces the observed polarization rotation. In the usual limit of large detuning and low saturation the atomic response is linear, and the total field radiated by the dipoles in an optical lattice is simply the sum of the forward scattered probe field and the Bragg scattered fields from each of the other lattice beams. A particularly simple situation occurs in our 1D lin-($\theta = 0$)-lin lattice, where the polarization is everywhere linear. When the lattice is detuned below atomic resonance ($\Delta < 0$) the atoms are confined near the antinodes of the standing wave, where the electric field strength and dipole moment is twice that for a single beam. This leads to twice the radiated field, and therefore twice the rotation of the probe polarization and twice the signal predicted by eq. 4. If, on the other hand, the lattice is detuned above atomic resonance, ($\Delta > 0$), atoms are confined near the nodes of the standing wave where the field strength and dipole moment vanishes, and the probe polarization is not rotated at all. Note that the total rate of photon scattering in the first case is increased by a factor of four over that of a single lattice beam, and in the second case decreased to zero, so that eqs. 8 and later still hold as long as we use the appropriate $\tau_s$. In practice the atomic localization is less than perfect, leading to a suppression of Bragg scattering by a factor $\beta = \exp[-\Delta k^2 \Delta z^2]$ (the Debye-Waller factor), where $\Delta k = 2k$ is the change in photon momentum and $\Delta z$ is the $1/e$ width of the atomic wavepackets [13]. Thus, for $\Delta < 0$ the Faraday signal is increased by a factor $1 + \beta \leq 2$, while for $\Delta > 0$ it is decreased by a factor $1 - \beta \geq 0$. A further suppression of Bragg scattering due to the non-isotropic dipole radiation pattern may also occur for some geometries, but in our case this correction is negligible.

Interestingly enough, this analysis suggests that Faraday rotation can be turned into a continuous probe of atom position in the optical lattice. Consider a situation in which the atomic sample has been optically pumped so that $\langle F_z \rangle = F$, and where the spins are kept aligned by a bias magnetic field along the z-axis. If $x$ denotes the position of an atom along the lattice axis, with $x = 0$ corresponding to an antinode of the standing wave, then the "signal" field is proportional to $1 + \exp(-i2kx)$, where the first term corresponds to forward scattering and the second term to Bragg scattering. The differential power $\Delta P_S$ detected by the polarization analyzer is then proportional to $\text{Re}[1 + \exp(-i2kx)] = 1 + \cos(2kx)$, which varies from 0 to 2 as the atom moves from a node to an antinode in the lattice. This implies that we can detect movement from a node to an antinode with the same SNR ratio as Larmor precession from $\langle F_z \rangle = 0$ to $\langle F_z \rangle = F$. Note that $\Delta P_s$ is an even function of $x$, i. e. we get information only about the distance the atom is displaced from the nearest node (or antinode). Of course a real atomic wavepacket contains a spread in $x$ which must be properly averaged over. It is precisely this average over a wavepacket trapped at the lattice nodes or antinodes that lead to the modification of the Faraday signal by the factors $1 \pm \beta$ as discussed above. In a more general situation both the internal and center-of-mass degreed of freedom will evolve over time, and both will be reflected in the Faraday signal. Furthermore, in more complex lattices (the simplest example of which are the 1D lin-$\theta \neq 0$-lin family) both the lattice intensity and polarization changes as a function of position, and a careful analysis is required to extract information about the overall dynamics.

**3. Experiment**

We evaluate the performance of our Faraday spectroscopy setup by observing Larmor precession of a sample of laser cooled and trapped Cesium atoms, and by comparing the measured signals to the predictions of section 2. Our basic experiment is similar to that of Isayama et al. [2], the main difference being that we use an optical lattice for both atom trapping and probing. Atoms are first collected and cooled in a standard vapor-cell magneto-optic trap (MOT), then cooled further in 3D optical molasses and a 1D near-resonance optical lattice, and finally transferred adiabatically to a 1D lin-$(\theta=0)$-lin far-off-resonance lattice tuned 10-100 GHz above or below the Cs D2 transition at 852 nm. The far-off-resonance lattice is formed by a MOPA semiconductor laser source with a total output power of ~0.4 W, which is split into a pair of lattice beams having a roughly Gaussian intensity distribution with $1/e$ radius of ~750 μm. This is about twice the typical radius $L = 350$ μm of our atom cloud, so that the lattice intensity is reasonably uniform across the atomic sample. We typically are able to load the lattice with a number of atoms in the range $N \sim 10^6 - 10^8$. Once trapped in the far-off-resonance lattice the sample is optically pumped within the $F = 4$ ground hyperfine manifold, using circularly polarized light and a bias magnetic field along $\varepsilon_y$ to produce a spin-coherent state aligned along the $y$-axis. We can measure the kinetic temperature of the atoms at this stage, from which we infer vibrational excitations in the range $\bar{n} \geq 0.2$ in the lattice microtraps. To initiate Larmor precession we quickly turn off (switching time ~ 5 μs) the bias field along $\varepsilon_y$, and apply a field of ~30 mGauss along $\varepsilon_x$.

A key aspect of our setup is the use of one lattice beam as a probe. As evident from eq. 10, the SNR in a measurement of Larmor precession is tied directly to the rate of decoherence arising from the scattering of probe photons. In our setup we avoid the introduction of extra decoherence that would occur if we used separate probe and lattice fields, as well as any disturbance of the atomic dynamics due to extra light-induced forces from the probe. To allow the observation of Faraday rotation in one of the lattice beams is relatively straightforward, requiring only a few minor additions to the standard 1D lattice setup as illustrated in fig. 1. First, the lattice beams are aligned at an angle ~12° away from counterpropagating. Second, a lens is inserted in the probe lattice beam to image the plane of intersection with the atom cloud at an aperture located ~1 m downstream. The aperture in the image plane allows us to select only the part of the lattice beam that has undergone Faraday rotation; the remainder which carries no signal but will contribute to the overall noise, is blocked. During measurements the aperture size is adjusted to optimize the SNR as discussed in section 2. Finally, the Faraday rotation signal is measured with a simple polarimeter consisting of a polarization beamsplitter cube and a differential photo-detector. The overall detection efficiency of the system is $\kappa \approx 0.29$.

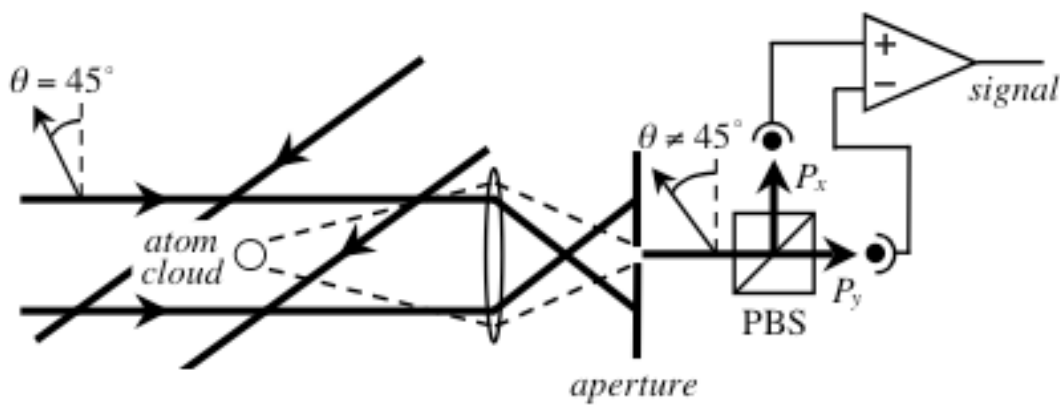

Figure 1. Setup for Faraday spectroscopy on atom samples in optical lattices. One of the linearly polarized lattice beams serves also as a probe beam, while an imaging lens and aperture ensures that only the part of the beam passing through the sample is detected. The polarization is analyzed by a simple polarimeter consisting of a polarization beamsplitter cube (PBS) and differential photo detector.

In order to achieve close to shot-noise limited detection care must be taken in the design and operation of the differential photo-detector. We use an auto-balancing circuit developed by Hobbs [14], which in principle allows very high rejection of common-mode noise arising from laser power fluctuations. The cancellation of common-mode noise is quite sensitive to displacement of the lattice beam due to mechanical vibration, a problem which we minimize by using large (~1 cm diameter) photodiodes. The autobalancing circuit is by design AC coupled, with a low frequency cut-off in the range 10-100 Hz and a high frequency roll-off at approximately 22 kHz. Further shaping of the detection frequency response was performed with a tunable 4th order bandpass filter. The complete system is essentially shot-noise limited from ~100 Hz – 20 kHz, apart from a small number of narrow noise peaks in the range below 2 kHz which carry too little power to be of significance, and which we ascribe to mechanical vibrations of the optics in the lattice beam path. Below ~50 Hz the noise spectrum is well above shot-noise, containing many broad peaks that seem to be associated with building vibrations which are not attenuated by the rigid legs of our optical table. This noise is effectively removed by choosing a low-frequency cut-off of ~2 kHz, which is still well below the Larmor frequency of ~10 kHz used in our experiment.

**3.1 Results**

We have performed Faraday spectroscopy on samples of Larmor precessing atoms for a wide range of optical lattice intensities and detunings. Figure 2 shows two typical signals, both real-time and averaged, taken with similar lattice depths and detunings $\Delta = \pm 50 GHz$, and with samples containing identical numbers of atoms in identical volumes. For negative detuning the SNR easily allows the observation of Larmor precession in real time. In the case of positive detuning the observed signal and SNR is smaller by a factor 4.7. Following the discussion in section 2.2, this corresponds to a Debye-Waller factor of $\beta \approx 0.65$, a value which varies slightly with lattice parameters but can be regarded as typical of our data. As a consistency check we note that the sign of the Faraday rotation angle is reversed from positive to negative detuning, as one would expect from the sign change of the real part of the polarizability when going from below to above resonance.

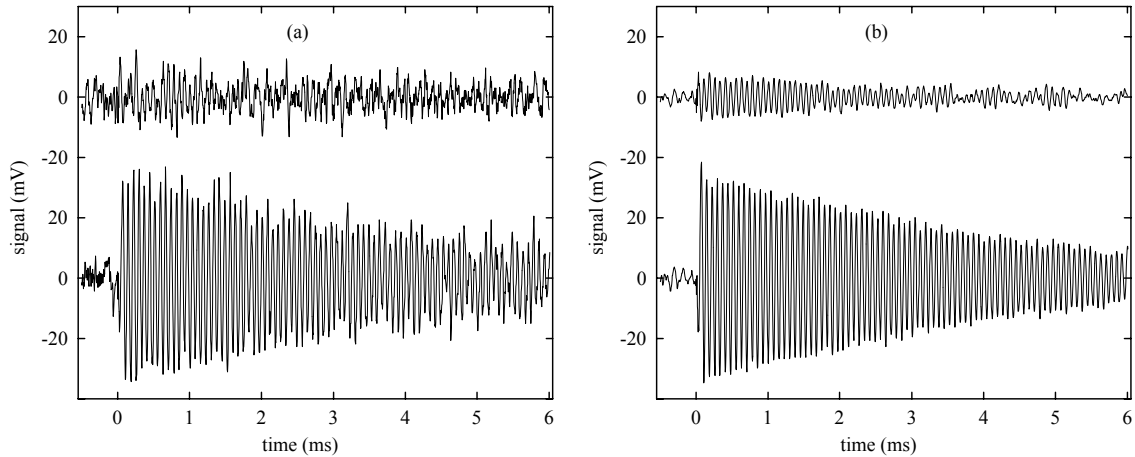

Figure 2. Time-dependent Faraday rotation, measured for a sample of a few $\times 10^6$ atoms which are spin-polarized and undergoing Larmor precession in a plane orthogonal to the probe beam. (a) Real-time Larmor precession signals, for $\Delta = 50$ GHz (top) and $\Delta = -50$ GHz (bottom). (b) Signal averages, for $\Delta = 50$ GHz (top) and $\Delta = -50$ GHz (bottom). For negative detuning the photon scattering time is $\tau_s = 6.2$ ms.

Experimental signal amplitudes were extracted from data such as that shown in fig. 2, by fitting the averaged signal with a damped sinusoid, $s(t) = A\exp(-t/\tau)\sin(\omega_L t + \varphi)$. Except for an initial transient caused by our 4th order bandpass filter, the fits are generally good and show only minor deviation from exponential damping. The root-mean-square (RMS) noise is estimated from the part of the real-time signal immediately preceding the start of Larmor precession, and the SNR computed by dividing the signal amplitude with the RMS noise. Note that this measure does not take into account noise intermodulation, i. e. excess noise on the signal at non-zero level due to fluctuations of the probe power. Since the total probe power is stable to $\leq 2\%$ we expect noise intermodulation to be significant only at signal levels much larger than those of fig. 2. Figure 3 shows the variation of the measured SNR as a function of $\tau_s$, the mean time between photon scattering events. Also shown is the prediction of eq. 10, for a total atom number $N \sim 1.7 \times 10^6$, which yields the best fit to our data. Figure 3 clearly displays the expected scaling over roughly two orders of magnitude in $\tau_s$ and one order of magnitude in SNR, and strongly supports the model developed in section 2. The absolute value of the observed SNR is more difficult to compare against theory, chiefly because it is difficult to measure the atom number with good accuracy. First we note that, in deriving a value for $N$ from the observed SNR, we have carefully taken into account several effects, including a Bragg scattering enhancement $1 + \beta < 2$, a small angle between the prepared spin state and the y-axis, and a small amount of birefringence in the beam path between the atoms and the polarimeter, which reduces the Faraday rotation signal by an overall factor of ~0.65. Secondly, we have measured $N$ more directly in two independent ways, one based on the amount of fluorescence emitted from the magneto-optic trap, and the other based on the amount of fluorescence emitted when the atom sample is released from the lattice and falls through a probe beam normally used for time-of-flight temperature measurements. In the first case we estimate $N \approx 1.5 \times 10^6$ and in the second case $N \approx 5 \times 10^6$, where the difference between these numbers may be regarded as representative of the absolute accuracy. Thus we conclude that the absolute SNR in our experiment is consistent with eq. 10, but that the experimental uncertainty on $N$ does not allow us to test the agreement to better than roughly a factor of three.

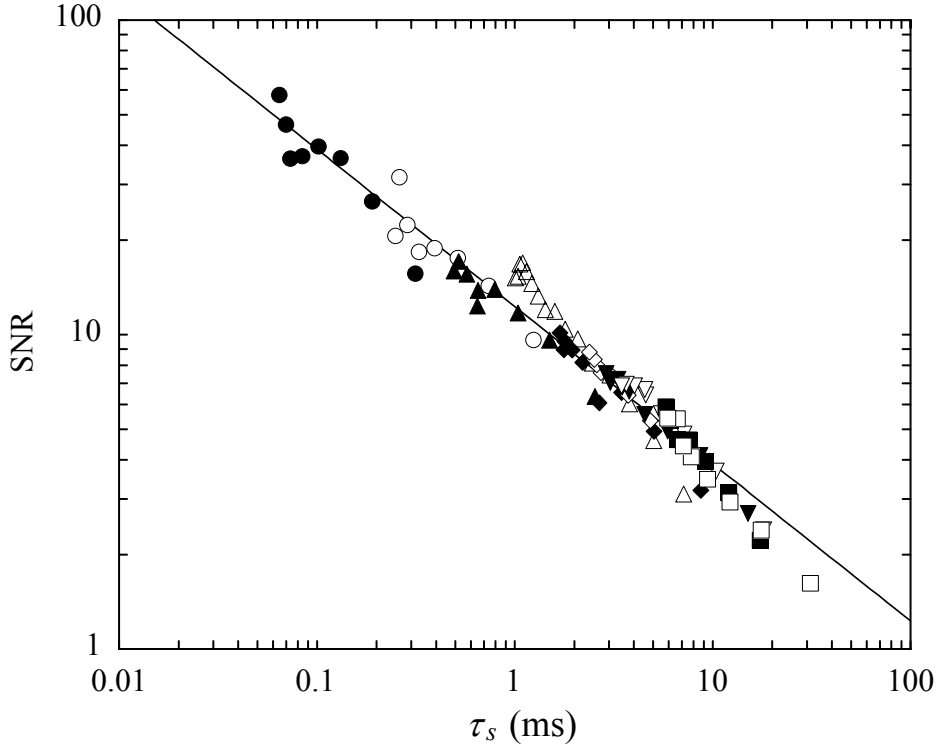

Figure 3. Signal-to-noise ratio (SNR) for Larmor precession, versus the mean time between photon scattering events ($\tau_s$). Symbols represent data taken at detunings of –9.9 GHz (●), -20.1 GHz (○), -30.2 GHz (▲), -40.2 GHz (△), -50.2 GHz (◆), -60.4 GHz (◇), -70.2 GHz(▼), -80.0 GHz (▽), -90.5 GHz (■), -100.0 GHz (□). Also shown is a best fit to the data, corresponding to an atom number $N = 1.7 \times 10^6$.

Ideally, depolarization of the atomic sample due to photon scattering will be the primary cause for decay of the Larmor precession signal. Figure 4 shows the observed damping time $\tau$ versus photon scattering time $\tau_s$. For scattering times below ~ 5 ms we find $\tau \approx \tau_s$, but for longer scattering times the observed damping times reach a plateau. The obvious conclusion is that some other mechanism limits the decay time for the Larmor precession signal to about ~ 5 ms. One likely cause is dephasing due to magnetic field variations, which only need to be few tens of μGauss across the sample in order to explain the observed limit.

In addition to the scaling of SNR with lattice parameters, it is also interesting to explore the maximum SNR which can be achieved with our setup. First, if we are interested only in a small bandwidth centered on the Larmor frequency, then much of the measurement noise can be eliminated with an appropriate bandpass filter. Second, our apparatus allows us to produce much larger samples, with as many as $10^8$ atoms in a cloud with $1/e$ radius $L \approx 750\,\mu m$. Figure 5 shows examples of two real-time signals acquired with such a sample, using detector bandwidths of 2 KHz (effective $\tau_{pd} = 125$ μs), and at detunings of $\sim -23$ GHz and $\sim -60$ GHz respectively. In either case the signal shape is virtually identical to those achieved with smaller samples, but the SNR has been greatly improved. Estimating the SNR as for the data in fig. 3 we find values (excluding intermodulation noise) of ~470 at −23 GHz, and ~250 at −60 GHz. These numbers are within a factor of two of those estimated using eq. 10 and the independently measured atom number, similar to the case for smaller atom samples.

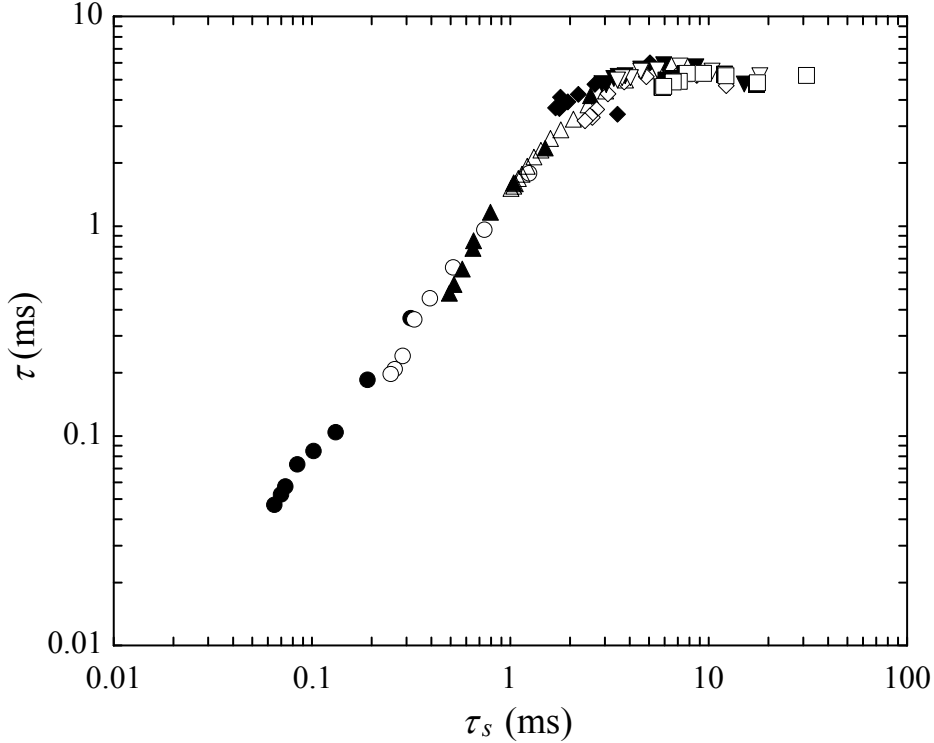

Figure 4. Dephasing times for the Larmor precession signal versus the mean time between photon scattering events. Symbols represent data taken at different detunings as in fig. 3.

With these larger atom samples it is relevant to consider the significance of quantum backaction. For the data in fig. 5a the on-resonance optical depth at the center of the atom cloud is $O \approx 6.6$ and the scattering time $\tau_s \approx 1.36$ ms. Thus we estimate a figure of merit of $\eta \approx 0.035$ in eq. 14, suggesting that we are still a factor ~30 away from the quantum limited regime. For the data in fig. 5b we are further away still, due to a slightly larger $\tau_s$ at this large detuning. To move closer to the quantum regime we could close down the detection aperture, choose probe/detector parameters $\tau_s \approx \tau$, and improve the detection system to avoid losses. In that case a sample with $N = 10^8$ and $L \approx 750$ μm has an optical depth $O \approx 10$ and gives a figure of merit $\eta \sim 0.6$, which is quite close to the quantum regime.

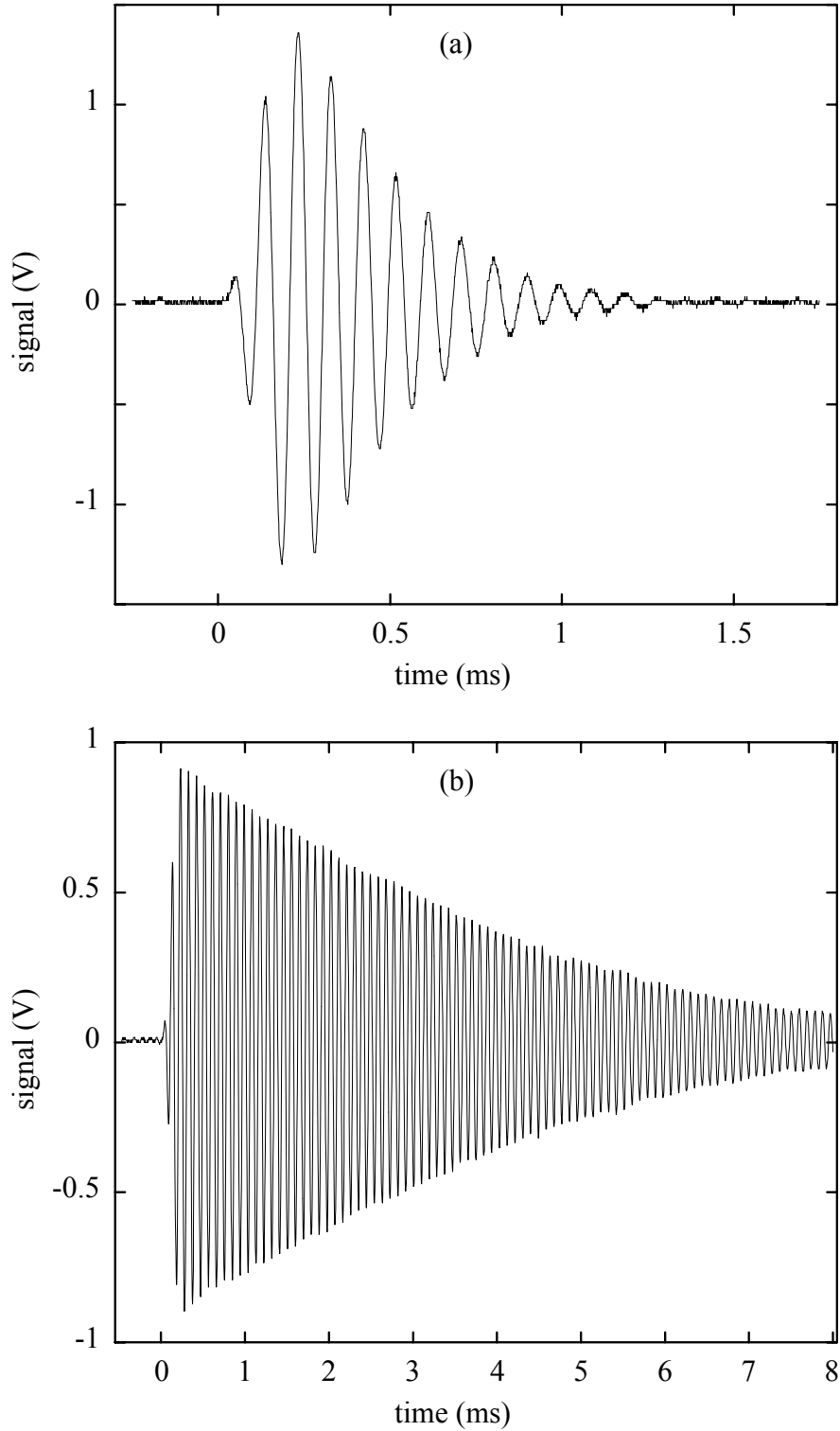

Figure 5. Real-time Larmor precession signal, measured with a sample of $\sim 10^8$ atoms. (a) Detuning of $-20$ GHz and $\tau_s = 8.0$ ms, (b) detuning of $-60$ GHz and $\tau_s = 1.4$ ms.

**4. Conclusion**

   We have developed a simple theory of Faraday spectroscopy in optical lattices, which predicts the sensitivity and SNR in measurements of the spin of laser cooled atomic samples. A key conclusion is that the SNR depends on probe and detector parameters solely through the ratio of the detector time constant to the mean time between photon scattering events. By comparing the measurement sensitivity against the spin projection noise we also determine approximately the conditions under which quantum backaction becomes significant. We further examine the implications of using one of the component beams of an optical lattice as probe, including the effect of Bragg scattering and the resulting dependence of the Faraday rotation on the atomic position in the lattice standing wave. We have carried out a demonstration experiment which observes Larmor precession of atoms trapped in a 1D optical lattice. Our data confirms the basic scaling of measurement sensitivity over nearly two orders of variation in the photon scattering rate, and the Larmor signal shows an absolute SNR which agrees well with theory within the experimental uncertainty. The data also clearly demonstrates the unavoidable tradeoff between sensitivity and decoherence due to photon scattering. With samples containing $\sim 10^8$ atoms we have achieved SNR's of a few hundred, though even in this case we remain at least a factor ~30 away from the regime of significant backaction.

   It is interesting to consider if our measurement can be extended into the backaction-limited regime. By matching the detector bandwidth and photon scattering rate, and by working with very large atomic samples, it seems quite plausible that this can be achieved. It will be much more challenging to realize a backaction-limited measurement with significant bandwidth margin, so as to allow the monitoring of coherent quantum dynamics and/or quantum feedback control. One possibility is to work with a Bose condensed sample, in which case the optical density on resonance can be as large as a few hundred for samples containing $\sim 10^6$ atoms. Another possibility is to perform Faraday spectroscopy in a buildup cavity [15]. A simple analysis suggests that the measurement sensitivity increases as the square root of the cavity finesse, so that substantial improvement can be achieved even with moderately high-finesse cavities. However, intracavity polarization spectroscopy brings with it new difficulties associated with the management of intracavity birefringence and Faraday activity, and it remains to be seen if this approach is feasible in practice.

   We thank I. H. Deutsch, B. P. Anderson, A. M. Steinberg and H. Mabuchi for helpful discussions and valuable suggestions. This work was supported by NSF grant no. PHY-0099582, by ARO grant no. DAAD19-00-1-0375, and by JSOP grant no. DAAD19-00-1-0359.